\documentclass[%
 reprint,
 amsmath,amssymb,
 aps,
]{revtex4-2}

\usepackage{graphicx}
\usepackage{dcolumn}
\usepackage{bm}

\usepackage{amsmath,scalerel,amssymb}
\usepackage{graphicx,stmaryrd}
\usepackage{dcolumn}
\usepackage{bm}
\usepackage{color}
\usepackage[breaklinks]{hyperref}

\usepackage{graphicx}
\newcommand{\intprod}{\mathbin{\raisebox{\depth}{\scalebox{1}[-1]{$\lnot$}}}}

\usepackage{sansmath}
\newcommand{\cO}{\mathcal O}

\newcommand{\Sbold}{\boldsymbol{S}}

\definecolor{darkgreen}{RGB}{9, 120, 9}
\definecolor{darkred}{RGB}{120, 9, 9}

\newcommand{\hatSigma}{\hat{\Sigma}}

\newcommand{\eps}{\epsilon}
\newcommand{\tK}{\mathtt{K}}

\newcommand{\PS}{\mathbb{PS}}

\newcommand{\ZZ}{\mathbb{Z}}

\newcommand{\Vol}{\text{Vol}}

\newcommand{\Ocal}{\mathcal{O}}

\newcommand{\Abold}{\boldsymbol{A}}

\newcommand{\cL}{\mathcal{L}}

\newcommand{\PT}{\mathbb{PT}}

\newcommand{\PPb}{\overline{\mathbb{P}}}
\newcommand{\CC}{\mathbb{C}}

\newcommand{\MM}{\mathbb{M}}
\newcommand{\Mcal}{\mathcal{M}}

\newcommand{\PP}{\mathbb{P}}

\newcommand{\pvec}{\boldsymbol{p}}

\newcommand{\AB}{\mathbb{A}}

\newcommand{\Zcal}{\mathcal{Z}}

\newcommand{\Ccal}{\mathcal{C}}

\newcommand{\PTc}{\mathcal{PT}}

\newcommand{\Tr}{\text{Tr}}

 \def\one{\mbox{1 \kern-.59em {\rm l}}}

 \def\C{{\mathbb C}}

   \def\cM{{\cal M}}

\begin{document}


\title{Toward a twistor action for chiral higher-spin gravity}

\author{Tung Tran}%
\affiliation{\vspace{4pt}
 Service de Physique de l’Univers, Champs et Gravitation,\\
Université de Mons, 20 place du Parc, 7000 Mons, Belgium
  \vspace{6pt}
}%

\begin{abstract}
A covariant twistor action for chiral higher-spin theory in (A)dS and flat space is constructed in term of a holomorphic Chern-Simons theory on twistor space. The action reproduces all known cubic vertices of chiral higher-spin theory in flat space. The spacetime action of the holomorphic Chern-Simons theory in flat space is also obtained.
\end{abstract}

\maketitle


\section{Introduction}\label{sec:1}
The idea of constructing viable interacting higher-spin theories that can avoid No-go theorems/results in flat space \cite{Weinberg:1964ew,Coleman:1967ad,Benincasa:2007xk} and AdS \cite{Maldacena:2011jn,Bekaert:2015tva,Sleight:2017pcz} has been going on for the past few decades. However, only a few higher-spin models can overcome the barriers posed by the No-go theorems \footnote{An up-to-date report on the development of higher-spin theories can be found in \cite{Bekaert:2022poo}.}. They are the $3d$ topological higher-spin theories \cite{Blencowe:1988gj,Bergshoeff:1989ns,Pope:1989vj,Fradkin:1989xt,Campoleoni:2010zq,Henneaux:2010xg,Grigoriev:2019xmp,Grigoriev:2020lzu}, conformal higher-spin gravity \cite{Segal:2002gd,Tseytlin:2002gz,Bekaert:2010ky,Haehnel:2016mlb,Adamo:2016ple}, chiral higher-spin gravity (chiral HSGRA)  \cite{Metsaev:1991mt,Metsaev:1991nb,Ponomarev:2016lrm,Metsaev:2018xip,Skvortsov:2018uru,Sharapov:2022faa,Sharapov:2022awp} and its contractions \cite{Ponomarev:2017nrr,Krasnov:2021nsq,Tran:2021ukl}, as well as higher-spin theories induced by the IKKT-matrix model \cite{Steinacker:2016vgf,Sperling:2017gmy,Sperling:2019xar,Steinacker:2022jjv}. In order to retain locality, which is one of the crucial features of field theories, one often has to give up unitarity. As a consequence, local higher-spin theories with propagating
degrees of freedom tend to be ``chiral'' in nature. It is,
therefore, not surprising that twistor theory \cite{Penrose:1967wn} is one of ideal
frameworks for constructing local higher-spin theories. See e.g. \cite{mason1996integrability} for an expedition in this direction. We note, however, that the chiral higher-spin theories are consistent truncations of some hypothetical complete theories which are unitary but are usually non-local. 

There are two main reasons why chiral HSGRAs can avoid No-go theorems/results. Firstly, chiral theories are higher-spin extensions or at least closed cousins of self-dual theories \cite{Chalmers:1996rq,Siegel:1992wd,Krasnov:2021cva}. Note that higher-spin extensions of self-dual Yang-Mills (SDYM), and self-dual gravity (SDGR) theories have been obtained recently in \cite{Ponomarev:2017nrr,Krasnov:2021nsq}. Secondly, constraints from higher-spin symmetry force all possible interactions to cancel each other out in the physical amplitudes in flat space. For instance, both conformal higher-spin gravity and chiral HSGRA have been shown to have vanishing tree-level amplitudes  \cite{Joung:2015eny,Beccaria:2016syk,Roiban:2017iqg,Ponomarev:2016cwi,Skvortsov:2018jea,Skvortsov:2020wtf,Skvortsov:2020gpn}. Furthermore, the chiral HSGRA was shown to be UV-finite at one-loop \cite{Skvortsov:2018jea,Skvortsov:2020wtf,Skvortsov:2020gpn}. 


In this work, we propose a covariant twistor action for the chiral HSGRA in (A)dS in response to the quest of covariantising the light-cone action of chiral HSGRA. Note that the proposed twistor action for chiral HSGRA is not fully determined as we do not know the explicit expression of what we call $S_c$ -- the correction accounts for the higher-spin diffeomorphism of twistor coordinates. However, the twistor Chern-Simons action allows us to compute all three-point scattering amplitudes of chiral HSGRA. We also present the spacetime action of the holomorphic Chern-Simons theory in flat space. As a consistency check, we find that its cubic vertices, after projecting to the light-cone gauge, match with the ones in \cite{Bengtsson:1983pd,Bengtsson:1986kh}. This partly solves a long-standing problem between the incompatibility of cubic interactions between higher-spin fields in Fronsdal's approach \cite{Bekaert:2010hp,Fotopoulos:2010ay,Manvelyan:2010je,Boulanger:2015ova,Conde:2016izb} and the light-cone formalism \cite{Bengtsson:1983pd,Bengtsson:1986kh}. The twistor origin of chiral HSGRA indicates that it must be integrable and one-loop exact. 

\section{The twistor theory}\label{sec:2}
\subsection{Twistor geometry/correspondence} 
Let $\PTc$ be the twistor space associated with a conformally flat Euclidean spacetime $\cM$ with cosmological constant $\Lambda$
\begin{align}\label{eq:metric}
    ds^2=\frac{dx_{\mu}dx^{\mu}}{(1+\Lambda x^2)^2}=\Omega^2dx_{\mu}dx^{\mu}\,,\qquad \mu=1,2,3,4\,.
\end{align}
$\PTc$ is defined as an open subset of $\PP^3$  \footnote{See e.g. \cite{Huggett:1986fs,Jiang:2008xw,Adamo:2013cra,Adamo:2017qyl} for a comprehensive review on twistor theory.} 
\begin{align}\label{PT}
\PTc=\left\{\Zcal^{A}:=(\lambda^{\alpha},\mu^{\dot\alpha})\,\big|\,I_{AB}\Zcal^A\hat{\Zcal}^B\neq0\right\}\,.
\end{align}
Here, $\Zcal^A$ are homogeneous coordinates of $\PP^3$. On $\PTc$, there is a quaternionic
conjugation that maps $\Zcal^A$ to its dual twistor $\hat{\Zcal}^A=(\hat{\lambda}^{\alpha},\hat{\mu}^{\dot\alpha})$ where
\begin{align}\label{eq:hatoperator}
    \hat{\lambda}^{\alpha}=(-\overline{\lambda^1},\overline{\lambda^0})\,,\qquad \hat{\mu}^{\dot\alpha}=(-\overline{\mu^{\dot 1}},\overline{\mu^{\dot 0}})\,.
\end{align}
Note that $\overline{\lambda^1}$ is simply the complex conjugation of $\lambda^1$. Furthermore, $I_{AB}$ is known as the infinity twistor that specifies the conformal factor $\Omega$ in \eqref{eq:metric}. The infinity twistor is a skew bi-twistor satisfying \cite{Penrose:1972ia}
\begin{align}
    \frac{1}{2}I^{AB}\epsilon_{ABCD}=I_{CD}\,,\qquad I_{AC}I^{BC}=\Lambda \delta_A{}^B\,.
\end{align}
The infinity twistor has the following representatives:
\begin{align}
    I^{AB}=\begin{pmatrix} \Lambda \epsilon^{\alpha\beta} &0\\
    0 &\epsilon^{\dot\alpha\dot\beta}
    \end{pmatrix}\,,\quad I_{AB}=\begin{pmatrix}  \epsilon_{\alpha\beta} &0\\
    0 &\Lambda\epsilon_{\dot\alpha\dot\beta}
    \end{pmatrix}\,,
\end{align}
which induces a Poisson structure \cite{Penrose:1976js,Ward:1980am} 
\begin{align}
    \widetilde{\Pi}=I^{AB}\overleftarrow{\partial_A}\wedge \overrightarrow{\partial_B}=\Lambda \frac{\overleftarrow{\partial}}{\partial \lambda^{\alpha}}\wedge \frac{\overrightarrow{\partial}}{\partial \lambda_{\alpha}}+\frac{\overleftarrow{\partial}}{\partial \mu^{\dot\alpha}}\wedge \frac{\overrightarrow{\partial}}{\partial \mu_{\dot\alpha}}\,,
\end{align}
on $\PTc$. The above Poisson structure also induces the following star-product on $\PTc$ \cite{Haehnel:2016mlb,Adamo:2016ple}
\begin{align}
    f\star g&:=fe^{\ell_p \widetilde\Pi}\wedge g=\sum_{k=0}^{\infty}\frac{\ell_p^k}{k!}f\,\widetilde\Pi^k\,g\,,\label{eq:starproduct}
\end{align}
where $\ell_p$ is some natural length scale that plays the role of a deformation parameter. At $k=1$, we recover the standard Poisson bracket
\begin{align}
    \{f,g\}_{\PTc}=f \,\widetilde \Pi \,g\,.
\end{align}
Note that we will sometime suppress the $\wedge$-products to shorten our expressions.

\subsection{The twistor action}\label{subsectwistor}
In constructing the twistor action for chiral HSGRA, it is useful to define the following Euler operator \cite{Haehnel:2016mlb,Adamo:2016ple}
\begin{align}\label{Euler1}
    \hatSigma=\Zcal^A\frac{\partial}{\partial \Zcal^A}\,,
\end{align}
to measure the weight in $\Zcal$ of any twistor expression. For instance, $\hatSigma\, D^3\Zcal=4$, where
\begin{align}
    D^3\Zcal=\epsilon_{ABCD}\Zcal^Ad\Zcal^B\wedge d\Zcal^C\wedge d\Zcal^D\,
\end{align}
is the canonical measure on $\PTc$. Then, our proposed twistor action for chiral HSGRA in (anti)de-Sitter space is 
\begin{align}\label{eq:Chern-Simonsaction}
\begin{split}
    \Sbold[\AB]&=S_{hCS}+S_c=\int D^3\Zcal\,L[\AB]+S_c\,.
    \end{split}
\end{align}
Here, \small
\begin{align}\label{eq:Chern-SimonsLagrangian}
\begin{split}
    L[\AB]&=\text{Tr}\big[\AB\star\AB+\frac{2}{3}\AB\star\AB\star\AB\big]\\
    &=\text{Tr}\big[\sum_{h\in \ZZ}\AB_{-h} \star \bar{\partial}\AB_h + \frac{2}{3}\sum_{h_i\in \ZZ}\AB_{h_1} \star \AB_{h_2} \star \AB_{h_3}\big]\,
    \end{split}
\end{align}
\normalsize
is the Lagrangian that obeys the constraint 
\begin{align}\label{Lconstraint}
    \hatSigma \,L[\AB]=-4\,,
\end{align}
so that we have a well-defined integral on $\PTc$ \footnote{See also \cite{Bittleston:2020hfv,Costello:2021bah} for previous work on holomorphic Chern-Simons action on twistor space for lower spin cases.}. Note that on $\PTc$
\begin{align}
   \bar{\partial}=d\hat Z^A\frac{\partial}{\partial \hat Z^A}\,,\quad \text{and}\quad  \AB=\sum_{h\in \mathbb Z}\AB_h\,,
\end{align}    
where $\AB_h\in\Omega^{0,1}(\PTc,\text{End}(E)\otimes\Ocal(2h-2))\,$ is a twistor field corresponding to a spacetime matrix-valued higher-spin fields of helicity $h$, and $E$ is some rank-$N$ vector bundle which is locally trivial on the restriction to any twistor line $X\subset\PTc$. 

Lastly, the term $S_c$ in  \eqref{eq:Chern-Simonsaction} is the correction to the holomorphic Chern-Simons action $S_{hCS}$ that accounts for higher-spin diffeomorphism of the coordinates $\Zcal^A$ \cite{Adamo:2013tja}:
\begin{align}\label{diffeo}
    \delta \Zcal^A=\sum_{h\in \ZZ}\{\Zcal^A,\xi_h\}\,,\qquad \xi_h\in \Gamma(\PTc,\Ocal(2h-2))
\end{align}
which results in a non-gauge-invariant measure \footnote{In this work, we do not have a proposal for $S_c$ but we expect its explicit form will appear elsewhere}. 
\section{Scattering amplitudes}\label{sec:3}
All 3-point tree-level amplitudes of chiral HSGRA in (A)dS can be computed as follows. By doing integration by parts, we observe that $I^{AB}\partial_A \partial_B f=0$. Hence, we can reduce the number of the star-products by one in each term of the action $S_{hCS}$. Therefore, \eqref{eq:Chern-Simonsaction} can be cast into the following form
\small
\begin{align}
    \Sbold=\int\Tr\big[\sum_h\AB_{-h} \bar{\partial}\AB_h + \frac{2}{3}\sum_{h_i}\AB_{h_1}  \AB_{h_2} \star \AB_{h_3}\big]+S_c'\,
\end{align}
\normalsize
where $S_c'$ accounts for $S_c$ in \eqref{eq:Chern-Simonsaction} and the remnants of $\partial/\partial \Zcal^A$ when acting on the holomorphic measure $D^3\Zcal$. The twistor representative of the $(0,1)$-form connection $\AB_h$ on $\PTc$ is chosen to be \cite{Haehnel:2016mlb,Adamo:2016ple}
\begin{align}\label{planewavetwistor}
\begin{split}
    \AB_{h_i}&=\int_{\mathbb{C}} \frac{dt_i}{t_i^{2h_i-1}}\bar{\delta}^2(t_i\lambda-\lambda_{i})e^{t_i[\mu\tilde{\lambda}_i]}\,,
\end{split}
\end{align}
in terms of the on-shell four-momentum $k_i^{\alpha\dot\alpha}=\lambda_i^{\alpha}\tilde\lambda_i^{\dot\alpha}$, which is a null vector on the tangent space of $(A)dS_4$. Here,
\begin{align}
    \bar{\delta}(az-b)=\frac{1}{2\pi i}d\bar{z}\frac{\partial}{\partial \bar{z}}\Big(\frac{1}{az-b}\Big)
\end{align}
is a $(0,1)$-form holomorphic delta function \cite{Witten:2004cp}. Parametrizing $\lambda_{\alpha}=(1,z)$ and $\lambda_{\alpha}'=(b,a)$, we define 
\begin{align}
    \bar{\delta}(\langle\lambda\lambda'\rangle)=\frac{1}{2\pi i}d\overline{\lambda^{\dot\alpha}}\frac{\partial}{\partial \overline{\lambda^{\dot\alpha}}}\frac{1}{\langle \lambda\lambda'\rangle}\,.
\end{align}
Finally, the projective version of the holomorphic delta function is defined by \cite{Haehnel:2016mlb}
\begin{align}
    \bar{\delta}_m(\lambda,\lambda')=\Big[\frac{\langle \xi \lambda\rangle}{\langle \xi\lambda'\rangle}\Big]^m\bar{\delta}(\langle\lambda\lambda'\rangle)=\int_{\mathbb{C}}\frac{dt}{t^m}\bar{\delta}^2(t\lambda-\lambda'),
\end{align}
which explains the origin of the twistor representative \eqref{planewavetwistor}. Here, the degree of $t_i$ (or the helicity $h_i$) essentially defines the weight in $\lambda$ of $\AB_{h_i}$. Furthermore, the crucial difference between our setup and the setup in \cite{Haehnel:2016mlb,Adamo:2016ple} is that there is a scalar field corresponding to $h=0$, which is essential for quantum consistency of chiral HSGRA.

It was shown in \cite{Nagaraj:2018nxq} that the plane wave solutions for higher-spin fields have the same structures with the ones in flat space. This explains why we can use the momentum eigenstates  \eqref{planewavetwistor} in (A)dS.

A simple computation shows that 
\small
\begin{align}\label{eq:starim}
    \AB_{h_2}\star\AB_{h_3}\sim \frac{\ell_p^k}{k!}t_2^{d_{k,h_2}}t_3^{d_{k,h_3}}\Big([23]+\Lambda\left\langle \frac{\partial}{\partial \lambda_2}\frac{\partial}{\partial \lambda_3}\right\rangle\Big)^k\,,
\end{align}
\normalsize
where $d_{k,h_i}=k+1-2h_i$. Following the steps in \cite{Adamo:2016ple}, we rewrite $D^3\Zcal$ as $\frac{d^4\Zcal}{\Vol\, \C^*}$, which allows us to integrate out $\mu$ and $\lambda$ variables to obtain 4-dimensional momentum delta-function $\delta^4(P)$. Note that 
\begin{align}
    \begin{split}
   \left\langle \frac{\partial}{\partial \lambda_2}\frac{\partial}{\partial \lambda_3}\right\rangle \delta^4(P)&=-[23] \Box_P\delta^4(P)\,,\\
   \text{where}\quad  \Box_P&=\frac{1}{2}\frac{\partial}{\partial P^{\alpha\dot\alpha}}\frac{\partial}{\partial P_{\alpha\dot\alpha}}\,.
   \end{split}
\end{align}
Furthermore, we use the conventions where $\langle ab\rangle=a^{\alpha}b_{\alpha}$ and  $[ab]=a^{\dot\alpha}b_{\dot\alpha}$. From \eqref{Lconstraint}, we can read off the constraint between the number of derivatives in \eqref{eq:starim} and helicities of the external states:
\begin{align}\label{helicityconstraint}
    k=h_1+h_2+h_3-1\,.
\end{align}
Integrating over $\mu$, we obtain
\begin{equation}
    \begin{split}
        \cM^{\Lambda}_3&(h_1,h_2,h_3)=\frac{\ell_p^k}{k!}\int d^2\lambda dt_1dt_2dt_3t_1^{1-2h_1}t_2^{d_{k,h_2}}t_3^{d_{k,h_3}}\\
      &\left([23]+\Lambda\left\langle \frac{\partial}{\partial \lambda_2}\frac{\partial}{\partial\lambda_3}\right\rangle\right)^k\bar{\delta}^2(t_1\tilde\lambda_1+t_2\tilde\lambda_2+t_3\tilde\lambda_3)\\
       &\times\bar{\delta}^2(t_1\lambda-\lambda_1)\bar{\delta}^2(t_2\lambda-\lambda_2)\bar{\delta}^2(t_3\lambda-\lambda_3)\,.
    \end{split}
\end{equation}

The integrals over $\lambda$ and $t_i$ variables can be performed trivially as in \cite{Haehnel:2016mlb,Adamo:2016ple}. For instance, we can use $\Vol\, \CC^*$ to fix $t_1=1$. Then, the integration over $\lambda$ gives us delta functions on the support at 
\begin{align}
    t_2=\frac{\langle 23\rangle}{\langle 31\rangle}\,,\quad t_3=\frac{\langle 32\rangle}{\langle 12\rangle}\,.
\end{align}
From here, it is a simple computation to integrate over $t_2$ and $t_3$. After a few more steps of manipulating spinors using momentum conservation, we arrive at:
\begin{align}\label{3ptAdS}
    \begin{split}
    \cM_3^{\Lambda}(h_1,&h_2,h_3)=\frac{\big[\ell_p(1-\Lambda \Box_P)\big]^{h_1+h_2+h_3-1}}{\Gamma[h_1+h_2+h_3]}\delta^4(P)\\
    &\times[12]^{h_1+h_2-h_3}[23]^{h_2+h_3-h_1}[31]^{h_3+h_1-h_2}\,.
    \end{split}
\end{align}
Note that we do not need to fix the kinetic part of the above 3-pt amplitudes by symmetry as in
\cite{Nagaraj:2018nxq,Nagaraj:2019zmk}. Furthermore, in the flat limit where $\Lambda\rightarrow 0$, we obtain the standard $\overline{\text{MHV}}_3$ amplitudes, which enables us to read off the cubic coupling constants:
\begin{align}\label{eq:couplingconstants}
    \Ccal_{h_1,h_2,h_3}=\frac{\ell_p^{h_1+h_2+h_3-1}}{\Gamma[h_1+h_2+h_3]}\,,\quad h_1+h_2+h_3>0\,.
\end{align}
A nice feature about our twistor construction is that the coupling constant $\Ccal_{h_1,h_2,h_3}$ is built in, and there is no need to derive it dynamically as in \cite{Metsaev:1991mt,Metsaev:1991nb,Ponomarev:2016lrm}. 

Using the map between the spinor-helicity formalism and the light-cone formalism, see e.g. \cite{Chalmers:1998jb,Bengtsson:2016jfk} where
\begin{align}
    i]=2^{1/4}\binom{\bar{p}_i\beta_i^{-1/2}}{-\beta_i^{1/2}}\,,\qquad  i\rangle =2^{1/4}\binom{p_i\beta_i^{-1/2}}{-\beta_i^{1/2}}\,,
\end{align}
we can express the square and angle brackets as
\begin{align}
    [ij]=\sqrt{\frac{2}{\beta_i\beta_j}}\PPb_{ij}\,,\qquad \langle ij\rangle=\sqrt{\frac{2}{\beta_i\beta_j}}\PP_{ij}\,.
\end{align}
Note that $\PPb_{ij}=\bar p_i\beta_j-\bar p_j\beta_i$ with $\pvec_i=(\beta_i,p^-_i,p_i,\bar{p}_i)$ is the momentum of the external field that has helicity $h_i$. Then, we can show that the above 3-pt amplitudes in the flat limit reduce to
\begin{align}\label{cubiclightcone}
    \cM_3^{\Lambda\rightarrow 0}=\Ccal_{h_1,h_2,h_3}\frac{\PPb_{23}^{h_1+h_2+h_3}}{\beta_1^{h_1}\beta_2^{h_2}\beta_3^{h_3}}\,,
\end{align}
which are the correct cubic vertices obtained previously in \cite{Bengtsson:1983pd,Metsaev:1991mt,Metsaev:1991nb}. As a remark, it would be interesting to establish the map between \eqref{3ptAdS} and the cubic vertices in (A)dS found by Metsaev in \cite{Metsaev:2018xip}.

\section{Spacetime action in flat space}\label{sec:4}

In this section, we obtain the spacetime action of the holomorphic Chern-Simons action in flat space where $\MM:=\lim_{\Lambda\rightarrow 0}\Mcal$ from the action \eqref{eq:Chern-Simonsaction}. To simplify the problem, we assume that all deformations are sufficiently small so that they will not affect the complex structures on twistor space to avoid the complication that arises from Kodaira's theory, see details in \cite{Kodaira2005}. 

First of all, notice that the twistor space in this case reduces to the usual flat/undeformed twistor space where 
\begin{align}
    \PT=\left\{Z^A=(\lambda^{\alpha},\mu^{\dot\alpha}:=F^{\dot\alpha}(\lambda,x))\,\big|\,\lambda_{\alpha}\neq 0\right\}\,.
\end{align}
Here, $\lambda_{\alpha}$ are coordinates on the Riemann sphere $X\cong\PP^1$ base of the fibration $\pi:\PT\rightarrow \PP^1$, and $\mu^{\dot\alpha}$ up the fibers of the normal bundle  
\begin{align}
    N_{X}:=T(\PT)|_{\PP^1}/T(\PP^1)\simeq \Ocal(1)\oplus \Ocal(1)\,.
\end{align}

The correspondence between $\PT$ and $\MM$ is given by the incidence relations 
\begin{align}\label{increl}
    \mu^{\dot\alpha}=F^{\dot\alpha}(x,\lambda)=x^{\alpha\dot\alpha}\lambda_{\alpha}\,,
\end{align}
where $x$ are complexified spacetime coordinates \cite{kodaira1962theorem,kodaira1963stability}. The inverse of the above reads
\begin{align}
x^{\alpha\dot\alpha}=\frac{\lambda^{\alpha}\hat{\mu}^{\dot\alpha}-\hat{\lambda}^{\alpha}\mu^{\dot\alpha}}{\langle \lambda \hat{\lambda}\rangle}\,.
\end{align}
Thus, each point $x\in\MM$ corresponds to a holomorphic, linearly embedded Riemann sphere $X\cong\PP^1\subset\PT$, and any point $Z\in\PT$ corresponds to a self-dual null $\alpha$-plane in $\MM$. 

It is convenient to define the following basis \cite{Mason:2005zm} on the corresponding space $\PS$, which is a projectivisation of undotted spinor bundle,
\begin{subequations}\label{eq:basis}
\begin{align}
    \bar{\partial}_0&=\langle \lambda\hat{\lambda}\rangle \lambda_{\alpha}\frac{\partial}{\partial \hat{\lambda}_{\alpha}}\,,\quad  &\bar{\partial}_{\dot\alpha}&=-\lambda^{\alpha}\partial_{\alpha\dot\alpha}\,,\\
    \bar{e}^0&=\frac{\langle \hat{\lambda}d\hat{\lambda}\rangle}{\langle \lambda \hat{\lambda}\rangle^2}\,,\qquad \qquad  &\bar{e}^{\dot\alpha}&=-\frac{\hat{\lambda}_{\alpha}dx^{\alpha\dot\alpha}}{\langle \lambda\hat{\lambda}\rangle}\,,
\end{align}
\end{subequations}
where $\bar{\partial}_{0}$ and $\bar{\partial}_{\dot\alpha}$ are $(0,1)$-vector fields, and $\bar{e}^{0}$, $\bar{e}^{\dot\alpha}$ are their dual $(0,1)$-forms, respectively. Note that the above basis can be defined according to the fact that $T(\PP^1)\cong \Ocal(2)$ and $T^*(\PP^1)\cong \Ocal(-2)$ with $\lambda$ being our reference of weight. Since,
\begin{align}\label{eq:flatcomplexstructure}
    \bar{\partial}:=\bar{e}^0\bar{\partial}_0+\bar{e}^{\dot\alpha}\bar{\partial}_{\dot\alpha}\quad \text{where}\quad  \bar{\partial}^2=0\,,
\end{align}
we will take $\bar{\partial}$ to be our definition of integrable complex structure on $\PS\cong \PP^1\times \MM$. Using \eqref{increl}, we can check that \begin{align}
    \bar{\partial}=d\hat\lambda^{\alpha}\frac{\partial}{\partial \hat\lambda^{\alpha}}+d\hat\mu^{\dot\alpha}\frac{\partial}{\partial \hat\mu^{\dot\alpha}}=d\hat Z^A\frac{\partial}{\partial \hat Z^A}\,,
\end{align}
which is the usual definition of the Dolbeault operator on $\PT$. Note that we will use the Dolbeault operator $\bar{\partial}$ to define twistor ``background''.

The analog of the Euler operator \eqref{Euler1} on $\PS$ is 
\begin{align}\label{Euler2}
    \hatSigma_{\lambda}=\lambda^{\alpha}\frac{\partial}{\partial \lambda^{\alpha}}\,,
\end{align}
where the constraint \eqref{Lconstraint} becomes
\begin{align}\label{LconstraintWoodhouse}
    \hatSigma_{\lambda}\,L[\AB]=-4\,.
\end{align}
For later convenience, we also define the following $(1,0)$-vector fields and their dual $(1,0)$-forms on $\PS$:
\begin{subequations}\label{eq:nicebasisfordeformation}
\begin{align}
   \partial_0&:=\frac{\hat{\lambda}_{\alpha}}{\langle \lambda \hat \lambda \rangle}\frac{\partial}{\partial \lambda_{\alpha}}\,, && &\partial_{\dot\alpha}&:=-\frac{\hat{\lambda}^{\alpha}}{\langle \lambda\hat\lambda\rangle}\partial_{\alpha\dot\alpha}\,,\\
   e^0&:=\langle \lambda d\lambda\rangle\,, && &e^{\dot\alpha}&:=\lambda_{\alpha}dx^{\alpha\dot\alpha}\,.
\end{align}
\end{subequations}
Here, $e^0$ is the holomorphic top-form of the fiber $\PP^1$. The following relations are useful
\begin{align}
    [\bar{\partial}_0,\partial_{\dot\alpha}]=\bar{\partial}_{\dot\alpha}\,,\qquad [\bar{\partial}_{\dot\alpha},\partial_0]=\partial_{\dot\alpha}\,.
\end{align}

To obtain spacetime action of the chiral HSGRA, it is more convenient to work on $\PS$ that has the following Poisson structure
\begin{align}\label{eq:holomorphicPoissonI}
    f\,\Pi\, g=\epsilon^{\dot\alpha\dot\beta}\partial_{\dot\alpha}f\wedge \partial_{\dot\beta}g
\end{align}
When $\partial_{\dot\alpha}$-vector field acting on any $(p,q)$-form, we must promote the above Poisson structure to
\begin{align}\label{eq:holomorphicPoisson}
    \omega\,\Pi\,\eta:=\{\omega,\eta\}_h=\eps^{\dot\alpha\dot\beta}\cL_{\partial_{\dot\alpha}} \omega \wedge \cL_{\partial_{\dot\beta}}\eta\,,
\end{align}
The holomorphic Poisson structure \eqref{eq:holomorphicPoisson} then induces 
\begin{align}\label{eq:astproduct}
    \omega \ast \eta :=\omega\, e^{\ell_p \Pi}\wedge \eta =\sum_{k=0}^{\infty}\frac{\ell_p^k}{k!}\omega \,\Pi^k\,\eta\,.
\end{align}
For simplicity, we will set $\ell_p=1$ from now, and require any $\omega\in \Omega^{p,q}(\PTc,\cO(n))$ to satisfy
\begin{align}
    \hatSigma_{\lambda}\intprod \omega=0\,,\qquad \qquad \cL_{\hatSigma_{\lambda}}\omega=n\,\omega\,.
\end{align}
Here, the notation $\hatSigma_{\lambda}\intprod \equiv \iota_{\hatSigma}$ is the interior product wrt. to the $\hatSigma_{\lambda}$ vector field. Furthermore, the Lie derivative $\cL_{\partial_{\dot\alpha}}$ acting on $\cO(n)$-valued $(p,q)$-forms can be defined via the Cartan's magic formula as
\begin{align}
    \cL_{\partial_{\dot\alpha}}\omega=\partial_{\dot\alpha}\intprod \eth \omega+\eth (\partial_{\dot\alpha}\intprod \omega)\,,
\end{align}
where we denote the usual exterior derivative on $\PS$ by $d_{\PS}\equiv \eth$. Note that $\eth$ is defined via a unique Chern connection on the bundle $\cO(n)\rightarrow\PP^1$ where \cite{Herfray:2016qvg}
\begin{align}
    \eth:=\partial+\bar{\partial}=d_{\mathbb S}+n\frac{\langle \hat \lambda d\lambda\rangle}{\langle \lambda \hat\lambda\rangle}\wedge\,.
\end{align}
Here,
\begin{align}
    d_{\mathbb S}:=e^0\partial_{0}+\bar e^0\bar{\partial}_0+dx^{\alpha\dot\alpha}\frac{\partial}{\partial x^{\alpha\dot\alpha}}
\end{align}
is the exterior derivative on the unprojective spinor bundle $\mathbb S$. 
It is easy to check that 
\begin{align}
    \eth e^0=\eth \bar{e}^0=0\,,\quad \eth e^{\dot\alpha}=e^0\wedge \bar{e}^{\dot\alpha}\,,\quad \eth \bar{e}^{\dot\alpha}=e^{\dot\alpha}\wedge \bar{e}^0\,.
\end{align}
This is the ``frame-dragging'' effect caused by the Lie derivative when it acts on vielbeins. In addition, since $\eth:=\partial+\bar{\partial}$, we get
\begin{align}
    \partial \bar{e}^0=0\,,\qquad \bar{e}^{\dot\alpha}=e^{\dot\alpha}\wedge \bar{e}^0\,.
\end{align}

Using the basis \eqref{eq:basis}, we decompose each twistor field as
\begin{align}
    \AB=\AB_0\bar{e}^0+\AB_{\dot\alpha}\bar{e}^{\dot\alpha}\,.
\end{align}
Furthermore, we have the following rules:
\begin{subequations}
\begin{align}
    \partial_0\intprod e^0&=1\,,\qquad \quad  \bar{\partial}_0\intprod\bar{e}^0=1\,,\\
    \partial_{\dot\alpha}\intprod e^{\dot\beta}&=\delta_{\dot\alpha}{}^{\dot\beta}\,,\quad \ \ \bar{\partial}_{\dot\alpha}\intprod\bar{e}^{\dot\beta}=\delta_{\dot\alpha}{}^{\dot\beta}\,.
\end{align}
\end{subequations}

Note that to reduce the number of $\ast$-product by one as in the case of holomorphic Chern-Simons action on $\PTc$, we need to assume that \cite{Sharma:2021pkl} \begin{align}
    \cL_{\partial_{\dot\alpha}}\AB^{\dot\alpha}=\partial_{\dot\alpha}\AB^{\dot\alpha}=0\,.
\end{align}
This will be our gauge condition. In this particular gauge, we can check that
\begin{align}
    \cL_{\partial_{\dot\alpha}}\cL_{\partial^{\dot\alpha}}\AB=0\,.
\end{align}
As a consequence, the origin Chern-Simons twistor action can be written as 
\small
\begin{align}\label{eq:actionCS}
    S[\AB]=\int_{\PS} \Tr\big[\sum_{h\in \mathbb{Z}}\AB_{-h} \bar{\partial}\AB_h + \frac{2}{3}\sum_{h_i\in \mathbb{Z}}\AB_{h_1} \AB_{h_2} \ast \AB_{h_3}\big]\,.
\end{align}
\normalsize
This is, indeed, a crucial fact since we want to have only one derivative, i.e. $\bar{\partial}$, in the kinetic term. Furthermore, we must have at least one positive-helicity field in \eqref{eq:actionCS} so that the constraint \eqref{LconstraintWoodhouse} can be implemented. 

A simple computation results in
\begin{align}
    \cL_{\partial_{\dot\alpha}}\AB=(\partial_{\dot\alpha}\AB_0+\AB_{\dot\alpha})\bar{e}^0+\partial_{\dot\alpha}\AB_{\dot\beta}\bar{e}^{\dot\beta}\,,
\end{align}
where $\partial_{\dot\alpha}\intprod\AB=0$. Then, it can be shown, albeit with some tedium, that
\begin{align}\label{eq:AAast}
    \begin{split}
    &\AB \ast \AB=\frac{1}{k!}[\partial_{\dot\alpha(k)}\AB_{\dot\beta},\partial^{\dot\alpha(k)}\AB_{\dot\gamma}]\bar{e}^{\dot\beta}\bar{e}^{\dot\gamma}\\
    &\qquad\  \, +\frac{1}{k!}[\partial_{\dot\alpha(k)}\AB_{0},\partial^{\dot\alpha(k)}\AB_{\dot\beta}]\bar{e}^0\bar{e}^{\dot\beta}\\
    &+\bar{e}^0 \bar{e}^{\dot\alpha}\frac{\partial_{\dot\gamma(k-1)}\AB_{\dot\gamma}\partial^{\dot\gamma(k)}\AB_{\dot\alpha}-\partial_{\dot\gamma(k)}\AB_{\dot\alpha}\partial^{\dot\gamma(k-1)}\AB^{\dot\gamma}}
    {(k-1)!}\,,
    \end{split}
\end{align}
where we used the convention 
$\partial_{\dot\alpha(k)}\equiv\partial_{\dot\alpha_1}...\partial_{\dot\alpha_k}$ to shorten our expressions.
\medskip

Unlike the case of SD HS-YM \cite{Tran:2021ukl}, where we can gauge fix $\AB_0\in \Omega^{0,1}(\PP^1,\cO(n)) = 0$ for $n\geq -1$ in Woodhouse gauge \cite{Woodhouse:1985id}. The situation here is significantly different since the deformation of twistor geometry is related to derivatives along the horizontal direction wrt. $T_x\MM$. In particular, there is an in-homogeneous contribution to the $\bar{e}^0\wedge \bar{e}^{\dot\alpha}$ component of the equation 
\begin{align}\label{eq:EOMastA}
    \bar\partial \AB+\AB\ast \AB=0\,,
\end{align}
From \eqref{eq:EOMastA}, we deduce that 
\begin{subequations}
\begin{align}
    0&=\bar\partial_0\AB_{\dot\alpha}-\bar{\partial}_{\dot\alpha}\AB_0+\frac{1}{2}\bar{\partial}_0\intprod\bar{\partial}_{\dot\alpha}\intprod\eqref{eq:AAast}\,,\label{eq:bare0eda}\\
    0&=\bar\partial_{\dot\alpha}\AB_{\dot\beta}+\frac{1}{k!}\partial_{\dot\gamma(k)}\AB_{\dot\alpha}\partial^{\dot\gamma(k)}\AB_{\dot\beta}\,.
\end{align}
\end{subequations}
Using the fact that $\bar{\partial}_0\partial_{\dot\alpha}=-\bar{\partial}_{\dot\alpha}$, we obtain the following solution for \eqref{eq:bare0eda}:
\begin{align}\label{eq:A0step1}
    \partial_{\dot\alpha}\AB_0=-\AB_{\dot\alpha}\,.
\end{align}
From \eqref{eq:A0step1}, we can further show that \begin{align}
    \bar{\partial}_0\AB_0=0\,,\qquad \bar{\partial}_0\AB_{\dot\alpha}=0\,.
\end{align}
Namely, $\AB_{\dot\alpha},\AB_0$ must be holomorphic in $\lambda$ when they have positive weight. We can now consider
\begin{align}\label{eq:generatingfunction}
    \begin{split}
    \AB_{h,\dot\alpha}\bar{e}^{\dot\alpha}=\lambda^{\alpha(2h-1)}A_{\alpha(2h-1),\dot\alpha}\,\bar{e}^{\dot\alpha}\,.
    \end{split}
\end{align}
This can be used to solve for the zero component of $\AB$ as:
\begin{subequations}
\begin{align}
    \AB_{h,0}^+\bar{e}^0&= -\frac{\bar{\partial}^{\dot\alpha}}{\Box} \AB_{\dot\alpha}\bar e^0\,,\qquad\quad  h>0\,,\label{eq:dangerousBox}\\
    \AB_{h,0}^-\bar{e}^0&=\frac{\hat{\lambda}_{\alpha(2|h|)}}{\langle \lambda \hat\lambda\rangle^{2|h|}}B^{\alpha(2|h|)}\bar{e}^0\,,\quad\ \   h\leq 0\,.
\end{align}
\end{subequations}
where $\lambda^{\alpha(s)}=\lambda^{(\alpha_1}...\lambda^{\alpha_s)}\,,\  \hat{\lambda}_{\alpha(s)}=\hat\lambda_{(\alpha_1}...\hat\lambda_{\alpha_s)}\,$. Note that while the non-local $\Box^{-1}$ may look `dangerous' at the moment, it will disappear after we integrating out all fibre coordinates (see below).
\medskip

To include the scalar field in the spectrum of chiral HSGRA, we can consider the following twistor field
\begin{align}
   \AB_{h=0^+}:= \vartheta=\frac{\hat{\lambda}_{\alpha}}{\langle \lambda\hat\lambda\rangle}\vartheta^{\alpha}{}_{\dot\alpha}\bar{e}^{\dot\alpha}\,.
\end{align}
Here, $\vartheta^{\alpha}{}_{\dot\alpha}$ is the auxiliary field associated with the scalar field, which can be integrated out by its own equation of motion as observed in \cite{Boels:2006ir}. It is not hard to show that
\begin{equation}\label{eq:varthetaexplicit}
    \begin{split}
    \vartheta^{\alpha}{}_{\dot\alpha}&=\Big(\partial^{\alpha}{}_{\dot\alpha}+\AB^{\alpha}{}_{\dot\alpha}\ast\Big)\AB_0\\
    &+\frac{1}{k!}\partial_{\dot\gamma_1\ldots \dot\gamma_{k+1}}\AB_{\dot\alpha}\partial^{(\dot\gamma_1}\ldots \partial^{\dot\gamma_k} \AB^{\dot\gamma_{k+1})}\,,
    \end{split}
\end{equation}
where $\AB_{\alpha\dot\alpha}\in\{ \oplus_s\Gamma(\PT,\text{End}(E)\otimes\Ocal(2s-2))\,|\,s\geq 1\}\,.$
The twistor action \eqref{eq:actionCS} reads
\small
\begin{align}\label{eq:beforeintegration}
    \begin{split}
    \Sbold&=\int_{\PS} \mho\Tr\Big[\AB_0(\bar{\partial}_{\dot\alpha}+\AB_{\dot\alpha}\ast )\AB^{\dot\alpha}+\frac{\lambda^{\gamma}\hat\lambda^{\beta}}{\langle \lambda\hat\lambda\rangle}\vartheta_{\alpha\dot\alpha}\,\vartheta_{\beta}{}^{\dot\alpha}\Big]+S_c\,,
    \end{split}
\end{align}
\normalsize
where the measure $\mho$ is \cite{Mason:2005zm}
\begin{align}
    \mho=D^3Z\,\bar{e}^0 [\bar{e}^{\dot\alpha} \bar{e}_{\dot\alpha}]=d^4x\frac{\langle \lambda d\lambda\rangle\wedge \langle \hat{\lambda}d\hat{\lambda}\rangle  }{\langle \lambda \hat{\lambda}\rangle^2}=d^4x\,\tK\,,
\end{align}
and $\tK$ is the top form on $\PP^1$. 
The spacetime action for chiral HSGRA on a flat background can be obtained by integrating out fibre coordinates using \cite{Woodhouse:1985id,Boels:2006ir,Jiang:2008xw}:
    \begin{align}\label{eq:bridge}
    \int_{\PP^1}\tK\, \frac{\hat{\lambda}_{\alpha(m)}\,\lambda^{\beta(m)}}{\langle \lambda \hat{\lambda}\rangle^{m}}=-\frac{2\pi i}{(m+1)}\epsilon^{\ \beta}_{ \alpha}...\epsilon^{\ \beta}_{ \alpha}\,,
\end{align}
where we adopted the same convention in \cite{Tran:2021ukl}. 

The resulting spacetime action (after substituting \eqref{eq:varthetaexplicit} to \eqref{eq:beforeintegration} and do some suitable rescaling) is the following action
\begin{align}\label{eq:spacetimeaction}
    \begin{split}
        S&= \langle  \boldsymbol{B}_{\alpha\alpha}| D^{\alpha}_{\ \dot\alpha} \boldsymbol{A}^{\alpha\dot\alpha}\rangle-\frac{1}{2}\langle D_{\alpha\dot\alpha} \widetilde{\Phi}|D^{\alpha\dot\alpha}\widetilde{\Phi}\rangle\\
        &+\langle \widetilde\Phi|[\![\boldsymbol A_{\dot\alpha},\boldsymbol A^{\dot\alpha}]\!]\rangle-\frac{1}{2}\langle [\![\Abold_{\dot\gamma},\partial^{\dot\gamma}\Abold_{\alpha\dot\alpha}]\!] | [\![\Abold_{\dot\beta},\partial^{\dot\beta}\Abold^{\alpha\dot\alpha}]\!]\rangle\,,
    \end{split}
\end{align}
where we define
\begin{align}\label{eq:bracketdef}
    \langle X|Y\rangle := \int dx^4 X_{\alpha(n)}Y^{\alpha(n)}\,.
\end{align}
To understand the above angled bracket notation, it is convenient to introduce
\small
\begin{subequations}\label{representative1}
\begin{align}
    \boldsymbol{A}^{\alpha\dot\alpha}&:=\sum_{s\geq 1}A^{\alpha(2s-1),\dot\alpha}y_{\alpha(2s-2)}\,,\quad y_{\alpha(n)}=y_{\alpha}\ldots y_{\alpha}\,\\
    \boldsymbol{A}^{\dot\alpha}&:=\sum_{s\geq 1}A^{\alpha(2s-1),\dot\alpha}y_{\alpha(2s-1)}\,,
\end{align}
\end{subequations}
\normalsize
as the generating function for positive-helicity fields, and
\begin{align}
    \boldsymbol{B}_{\alpha\alpha}:=\sum_{s\geq 1}B_{\alpha(2s)}\tilde y^{\alpha(2s-2)}\,,\quad \tilde y^{\alpha(n)}=\tilde y^{\alpha}\ldots\tilde y^{\alpha}\,,
\end{align}    
as generating functions for non-positive helicity fields. Lastly, the field $\widetilde \Phi$ contains both positive and negative helicity fields since its originated from $\AB_0$:
\begin{align}\label{representative2}
    \widetilde{\Phi}:=\sum_{s\geq 0}\frac{\partial^{\beta\dot\alpha}}{\Box}A^{\alpha(2s-1)}{}_{\dot\alpha}y_{\beta\alpha(2s-1)}+B_{\alpha(2s)}\tilde y^{\alpha(2s)}\,, 
\end{align}
where we note that the coefficients come with $y^{\alpha}$ are positive helicity fields and the coefficients come with $\tilde y^{\alpha}$ are negative helicity fields, and $B_0=\phi$ the scalar field.
\medskip

To proceed, we will treat the commuting auxiliary variables $y_{\alpha},\tilde y^{\alpha}$ as creation and annihilation oscillators with the property that each $\tilde y$ will consume one $y$ and give us a Kronecker delta for contraction. Therefore, at free level, the second term in \eqref{eq:spacetimeaction} reduces to the usual kinetic term of free fields in spacetime. Furthermore,
\begin{align}
    D^{\alpha}_{\ \dot\alpha}\bullet:=\partial^{\alpha}_{\ \dot\alpha}\bullet+[\![\boldsymbol{A}^{\alpha}_{\ \dot\alpha},\bullet]\!]\,,
\end{align}
where the double square bracket takes the following form by virtue of the $\ast$-product \eqref{eq:astproduct}, e.g.
\small
\begin{align}
    \begin{split}
    &[\![A^{\alpha(2s-1),}{}_{\dot \gamma},A^{\alpha(2s'-1),\dot\gamma}]\!]\\
    &:=\frac{1}{k!}[\partial_{\alpha\dot\beta_1}...\partial_{\alpha\dot\beta_k}A^{\alpha(2s-1),}{}_{ \dot\gamma},\partial_{\alpha}^{\ \dot\beta_1}...\partial_{\alpha}^{\ \dot\beta_k}A^{\alpha(2s'-1),\dot\gamma}]\,,
    \end{split}
\end{align}
\normalsize
where we recall that we have set $\ell_p=1$ in this section for simplicity. Here, all un-dotted indices of the partial derivatives are understood to contract with the ones of physical fields in every possible way. 
\medskip

To this end, let us explain how the $(+,+,+)$ cubic vertices come to be. Due to the contraction between derivatives originated from the $\ast$-product and the one in $\tilde\Phi_{h\geq 0}$, we can form $\Box$ to cancel out the non-local $\Box^{-1}$ in \eqref{eq:dangerousBox}. To illustrate, let us look at the term $\langle \widetilde \Phi|[\![\boldsymbol A_{\dot\alpha},\boldsymbol A^{\dot\alpha}]\!]\rangle$. Using the representatives \eqref{representative1} and \eqref{representative2}, we obtain
\small
\begin{align}\label{step1}
    \frac{\partial^{\bullet \dot\gamma}}{\Box}A^{\alpha(2s_1-1)}{}_{\dot\gamma}\big[\partial_{\alpha\dot\beta}\ldots \partial_{\alpha\dot\beta}A^{\alpha(2s_2-1)}{}_{\dot\circ},\partial_{\bullet}{}^{\dot\beta}\ldots \partial_{\alpha}{}^{\dot\beta}A^{\alpha(2s_3-1)\dot\circ}\big]
\end{align}
\normalsize
Upon integrating by part and applying the identity
$\partial^{\bullet\dot\gamma}\partial_{\bullet\dot\beta}\sim \Box \epsilon^{\dot\gamma}{}_{\dot\beta}$, we can cancel the $\Box^{-1}$ in \eqref{step1}. This leaves us with the following $(+,+,+)$ vertices:
\small
\begin{align}
    A^{\alpha(2s_1-1)}{}_{\dot\bullet}[\partial_{\alpha\dot\gamma}\ldots \partial_{\alpha\dot\gamma}A^{\alpha(2s_2-1)}{}_{\dot\alpha},\partial_{\alpha}{}^{\dot\gamma}\ldots\partial_{\alpha}{}^{\dot\gamma}\partial_{\alpha}{}^{\dot\bullet}A^{\alpha(2s_3-1),\dot\alpha}]\,.
\end{align}
\normalsize
Notice that there must be at least one extra pair of derivatives coming from the $\ast$-product to generate the $(+,+,+)$ vertices. In addition, the all-plus vertices have maximal number of derivatives allowed by kinematics. As such, $(+,+,+)$ vertices represent non-minimal couplings. 



\section{Discussion}\label{sec:5}
In this letter, we have constructed a covariant action for chiral HSGRA in (A)dS from a Chern-Simons action on twistor space. The twistor origin of chiral HSGRA indicates that it must be  integrable and one-loop exact. It is intriguing to ask whether we can have a world-sheet description for the chiral HSGRA to gain control over non-locality issues the moment we step outside the self-dual sectors. It will be a crucial step in finding a higher-spin theory with unitary completion where we have total control of non-local interactions. 
\medskip

One of the fundamental questions the results of this letter can address is whether having a covariant form for interacting higher-spin theories is an advantage. As in other covariant formulations of higher-spin theories, the twistor construction provides us a clear view about higher-spin symmetry that governs the chiral HSGRA. The covariant action of the chiral HSGRA also enable us to see some vertices in terms of spacetime derivatives that can not be obtained using Fronsdal's approach. However, note that the computation of scattering amplitudes of the action \eqref{eq:spacetimeaction} is more involved compared to the calculations in the light-cone gauge even at tree-level. 
\medskip

Contrary to the old folklore, recent developments in constructing higher-spin interactions by means of the light-front formalism or free differential algebra show that there exists a smooth deformation between the vertices in flat space and (A)dS \cite{Metsaev:2018xip,Sharapov:2022awp}. In this paper, we also confirm that such deformation exists at the level of the twistor action \eqref{eq:Chern-Simonsaction}.

\medskip

Lastly, chiral HSGRA is known to admit $U(N)$, $O(N)$ and $USp(N)$ gaugings \cite{Metsaev:1991mt,Metsaev:1991nb,Skvortsov:2020wtf}, see also \cite{Konstein:1989ij}. From this perspective, it is plausible that chiral HSGRA will become the self-dual part of the HS-IKKT matrix model with truncated higher-spin spectrum \cite{Steinacker:2016vgf,Sperling:2017gmy,Sperling:2019xar,Steinacker:2022jjv} in the deep quantum regime where spacetime coordinates no longer commute. We postpone the study of finding the connection between chiral HSGRA and HS-IKKT for future work.

\begin{acknowledgments}
The author is grateful for valuable discussions with Tim Adamo, Thomas Basile, Roland Bittleston, Yannick Herfray, Atul Sharma, Zhenya Skvortsov and Harold Steinacker. This work is partially supported by the Fonds de la Recherche Scientifique under Grants No. F.4503.20 (HighSpinSymm), Grant No. 40003607 (HigherSpinGraWave), T.0022.19 (Fundamental issues
in extended gravitational theories) and the funding from the European Research Council (ERC) under Grant No. 101002551.

\end{acknowledgments}
\bibliographystyle{apsrev4-1}
\bibliography{twistor.bib}
\end{document}